\documentclass[a4paper,12pt]{article}         
\usepackage{amsmath,amsfonts,amsthm,amssymb}  
\usepackage[utf8]{inputenc} 
\usepackage[T1]{fontenc}    
									   
\usepackage{graphicx,caption}            
\usepackage{color}                       
\usepackage{hyperref}                    
\hypersetup{breaklinks=true}

\usepackage{breakurl}

\usepackage{grffile}

\usepackage{makeidx}                  
\usepackage{acronym}
\usepackage{mathrsfs}
\usepackage{float}
\usepackage{vruler}                    
\usepackage{palatino, url, multicol}
\usepackage{setspace}                  
\usepackage{txfonts}

\usepackage{tikz}
\usepackage{tikz-3dplot}
\usetikzlibrary{shapes.geometric, arrows, calc, 3d}

\usepackage[numbers,sort&compress]{natbib}   
\usepackage{import}

\usepackage{caption}
\usepackage{subcaption}

\usepackage{listings}

\usepackage{comment}
\usepackage[title]{appendix}

\restylefloat{figure}

\title{Bridging Statistical Scattering and Aberration Theory: Ray Deflection Function - I: Theoretical Framework}

\author{Netzer Moriya}
\date{}

\begin{document}

\maketitle

\begin{abstract}
This paper introduces a new conceptual framework that recasts surface roughness effects as a "ray deflection function" (RDF)
which can be statistically represented through a modified Zernike-Fourier hybrid approach that directly connects the PSD with 
statistical aberration coefficients through spectral overlap integration. By establishing a direct mathematical 
relationship between
the power spectral density (PSD) of surface imperfections and the statistical distribution of aberration coefficients, we
develop a formalism that bridges known probabilistic scattering theory with deterministic aberration analysis.
This transformation allows surface roughness to be seamlessly integrated with other optical aberrations
by expressing its effects through equivalent modifications to the ideal mirror shape.
This framework provides computational advantages for ray-tracing simulations while maintaining statistical fidelity to
established scattering models, particularly for predicting the three-dimensional structure of imperfect focal bodies in
optical systems.
\end{abstract}

\section{Introduction}
Traditional treatments of surface roughness in optical systems often require either detailed physical modeling of 
surface topography or statistical scatter models that operate primarily in the angular domain. 

We propose an intermediate modeling approach that transforms the statistical properties of surface imperfections into a 
deterministic Ray Deflection Function (RDF), which can be applied globally to rays reflecting from an otherwise 
idealized optical surface.

We define the RDF, denoted by $\mathbf{D}(\mathbf{r})$, as the local angular deflection experienced by a ray upon reflection 
due to a height perturbation at surface point $\mathbf{r} = (x, y)$.
Assuming scalar wave optics and small perturbations, this deflection can be expressed in terms of the local phase gradient:
\begin{equation}
    \mathbf{D}(\mathbf{r}) = \frac{1}{k} \nabla \Phi(\mathbf{r}),
\end{equation}
where $\Phi(\mathbf{r})$ is the phase delay induced by surface height variations, and $k$ is proportional to the optical wavenumber
(see in what follows).

This relation is derived from the eikonal approximation and is valid under the following assumptions:

\begin{itemize}
    \item the surface perturbations are smooth and differentiable,
    \item the incident and reflected angles are small (small-angle or paraxial approximation),
    \item and the wavefront phase is well-described by scalar field theory (i.e., polarization and vector 
	diffraction are negligible).
\end{itemize}

Under these conditions, the RDF provides a deterministic and spatially resolved mapping from surface microstructure to 
ray angle deflection, enabling direct statistical analysis of angular scatter based on spatial features.

\vspace{0.4cm}

This approach offers several advantages:
\begin{itemize}
    \item Computational efficiency in ray-tracing simulations,
    \item A direct connection between surface roughness statistics and aberration theory,
    \item A framework for analyzing three-dimensional focal volume effects.
\end{itemize}

The key insight is that for many practical applications, the exact physical roughness pattern is less important than 
its statistical effect on the reflected light distribution. 
By creating a deterministic function that reproduces the statistical properties of surface roughness, 
we can greatly simplify the modeling while maintaining physical accuracy.

A central contribution of our approach is the transformation of statistical surface roughness into equivalent deterministic 
aberration terms that directly modify the ideal mirror shape.
We find that the magnitude of this effective shape change scales with the RMS roughness while preserving the statistical 
characteristics of the original PSD model. 
This equivalence between surface roughness and aberration terms represents a significant conceptual bridge between 
statistical scattering models and deterministic optical design approaches.

This paper focuses on establishing the mathematical and theoretical foundations of the ray deflection function approach. 
While we provide analytical validation of the framework, the emphasis here is on the rigorous development of the conceptual 
and mathematical basis that bridges statistical scattering models and deterministic aberration analysis. 
The theoretical framework presented here could serve as a foundation for numerical simulations and practical implementations 
in optical system modeling.

This paper presents the theoretical foundations of the Ray Deflection Function (RDF) approach. 
A companion paper \cite{moriya2025b} provides comprehensive numerical validation of this framework through detailed 
simulations and comparisons with established scattering models.

\section{Theoretical Foundation}

While our approach builds upon the Harvey-Shack formalism~\cite{harvey_shack} described in the following, it's worth noting that several other 
theoretical frameworks address surface scattering, including Beckmann-Spizzichino electromagnetic theory ~\cite{beckmann_spizzichino}, 
Rayleigh-Rice vector perturbation theory, and the Kirchhoff approximation \cite{Harvey2007}. These established approaches, 
like primarily characterize roughness effects through scattered intensity distributions or BRDFs in the angular domain. 
In contrast, our ray deflection function framework takes a fundamentally different approach by reformulating surface 
roughness as statistical aberration contributions. This represents a new conceptual bridge between the statistical 
treatment of surface scattering and deterministic aberration theory, allowing roughness to be integrated with other optical 
aberrations in a unified mathematical framework, rather than treated as a separate scattering phenomenon.

\subsection{Connection to Harvey–Shack Theory}
The Harvey–Shack theory shows that the bidirectional reflectance distribution function (BRDF) of a rough surface is directly proportional to the power spectral density (PSD) of its height variations~\cite{harvey_shack}:
\[
\mathrm{BRDF}(\alpha_s-\alpha_i,\;\beta_s-\beta_i)
\;\propto\;
\mathrm{PSD}\!\Bigl(\tfrac{\alpha_s-\alpha_i}{\lambda},\;\tfrac{\beta_s-\beta_i}{\lambda}\Bigr),
\]
where \(\alpha\) and \(\beta\) are the direction cosines of the incident (\(i\)) and scattered (\(s\)) rays.  
This relationship furnishes a rigorous link between surface statistics and angular scattering~\cite{beckmann_spizzichino}.

\subsection{Ray Deflection Function Formulation}
We now construct a \emph{ray deflection function} \(\mathbf D(\mathbf r_0)\) that, when added to the nominal reflection direction \(\mathbf r_0\), reproduces the same statistical scattering as the physical surface:
\begin{equation}
\mathbf r \;=\;\mathbf r_0 + \mathbf D(\mathbf r_0).
\label{eq:r_0}
\end{equation}

We show below that

\begin{equation}
\mathbf D(\mathbf r_0)
\;=\;-\,\frac{\lambda}{4\pi}\,\nabla\Phi(x,y),
\label{eq:Dr0}
\end{equation}

where $\mathbf{r_0} = \mathbf{r_0}(x,y)$ represents the position vector on the optical surface 
with Cartesian coordinates $(x,y)$, and $\nabla\Phi(x,y)$ is the gradient of the phase 
function, chosen to match the surface PSD, at that position.

\paragraph{Derivation.}  
Let \(h(x,y)\) be the local surface deviation and assume small incidence angle (\(\cos\theta_i\approx1\)).  The round‐trip optical path difference (OPD) is
\begin{equation}
\Delta\ell(x,y)
=2\,h(x,y)\cos\theta_i
\;\approx\;2\,h(x,y),
\end{equation}
so the reflected wavefront acquires phase
\begin{equation}
\Phi(x,y)
=\frac{2\pi}{\lambda}\,\Delta\ell
=\frac{4\pi}{\lambda}\,h(x,y).
\end{equation}
By Fermat’s principle and the eikonal approximation, a transverse phase gradient induces an angular tilt
\[
\delta\boldsymbol\theta
=-\,\frac{\lambda}{2\pi}\,\nabla\Phi(x,y),
\]
where the minus‐sign reflects deflection toward decreasing phase.  In a reflecting geometry this tilt \(\delta\boldsymbol\theta\) \emph{is} the ray deflection,
\[
\mathbf D
=\delta\boldsymbol\theta
=-\,\frac{\lambda}{2\pi}\,\nabla\Phi
=-\,\nabla\bigl(2\,h\bigr)
=-\,\nabla\,\Delta\ell.
\]
Noting that \(\Phi=(4\pi/\lambda)h\), one immediately recovers \eqref{eq:Dr0}.  
By synthesizing \(\Phi(x,y)\) with the same PSD as the true surface, equation~\eqref{eq:Dr0} applied to an ensemble 
of rays yields statistically equivalent scattering to the physical roughness.

For high-NA optics or grazing incidence systems, additional corrections may be required.

\subsubsection{Validity Range of Small-Angle Approximation}

The relationship between ray deflection and the phase function gradient assumes small angular deviations (\ref{eq:Dr0}), is 
derived from the first-order Taylor expansion of the 
reflection operation and is valid when the angular perturbations are sufficiently small. To quantify the validity range, 
we need to establish explicit criteria.

For a surface with RMS roughness $\sigma$ and correlation length $l_c$\footnote{The correlation length of an optical surface 
represents the lateral distance over which surface heights become statistically independent.}, the RMS slope is 
approximately $\sigma/l_c$. The resulting angular deviation in reflection is approximately twice the slope, leading to:

\begin{equation}
\theta_{\text{RMS}} \approx 2\frac{\sigma}{l_c}
\end{equation}

The small-angle approximation (where $\sin\theta \approx \theta$ and $\cos\theta \approx 1$) introduces an error of less 
than 1\% when $\theta < 0.1$ radians ($\approx 5.7$ degrees). For the approximation used in our ray deflection function, we require:

\begin{equation}
2\frac{\sigma}{l_c} < 0.1 \text{ radians}
\end{equation}

This yields the criterion:

\begin{equation}
\frac{\sigma}{l_c} < 0.05
\end{equation}

For typical optical surfaces with $\sigma \approx 10$ nm and $l_c \approx 1$ mm~\cite{duparre2002surface}, 
this gives $\sigma/l_c \approx 10^{-5}$, well within the validity range of the approximation. 
Even for rougher surfaces with $\sigma \approx 100$ nm and $l_c \approx 100$ $\mu $m, the ratio 
is $\sigma/l_c \approx 10^{-3}$, still satisfying the criterion.

For high-NA optics or grazing incidence systems where the angles of incidence or reflection approach 90 degrees, 
additional geometric corrections are required to account for the obliquity factor $\cos\theta_i$:

\begin{equation}
\mathbf{D}(\mathbf{r}_0) = -\frac{\lambda}{4\pi\cos\theta_i}\nabla\Phi(x,y)
\end{equation}

The validity range for these cases becomes more restrictive:

\begin{equation}
\frac{\sigma}{l_c} < 0.05\cos\theta_i
\end{equation}

This implies that for grazing incidence angles (e.g., $\theta_i = 80$ degrees), the roughness-to-correlation length ratio must be 
approximately ten times smaller ($\frac{\sigma}{l_c} < 0.0087$) to maintain accuracy.

These criteria ensure that our ray deflection function approach correctly models the statistical behavior of surface roughness effects within well-defined boundaries of applicability.

\subsubsection{Mathematical Derivation}

To derive the explicit form of the ray deflection function from a given PSD, we establish a rigorous connection 
through the phase function $\Phi(x,y)$. 

The ray deflection function $\mathbf{D}(\mathbf{r}_0)$ operates on the 
nominal reflection vector $\mathbf{r}_0$ from a smooth surface to produce a perturbed ray direction, as 
described in Equation~\ref{eq:r_0}. Under the 
small-angle approximation, this perturbation can be modeled as the gradient of a scalar phase function $\Phi(x,y)$, 
as shown in Equation~\ref{eq:Dr0}, where a path difference of $\lambda$ corresponds to a phase change of $2\pi$.

To ensure that the phase function $\Phi(x,y)$ reproduces the statistical properties of surface roughness (as described 
by the power spectral density, or PSD), we construct it using a basis expansion. For surfaces defined over circular 
apertures, Zernike polynomials $Z_j(\rho,\theta)$ provide an orthogonal and physically meaningful basis for 
representing wavefront aberrations. Accordingly, we express the phase as:
\begin{equation}
\Phi(x,y) = \sum_{j=1}^{N} C_j Z_j(\rho,\theta)
\label{PsumC}
\end{equation}

The coefficients $C_j$ are statistically assigned to match the target PSD (detailed in Section \ref{sec:DAT}), ensuring that the 
resulting ray deflection function generates a scattered ray distribution consistent with the underlying surface statistics. 
This process is described in more detail in Appendix~\ref{appendix:RDF}.

This formulation provides a direct link between the statistical properties of the surface roughness and the resulting ray deflections.

The phase function can be represented either using the Zernike polynomial~\cite{Dai1996} expansion as in \ref{PsumC} above,
or directly in terms of spatial frequency components~\cite{VanTreesBook2001}:

\begin{equation}
\Phi(x,y) = \sum_{p,q} A_{pq} \cos(2\pi f_p x + 2\pi f_q y + \phi_{pq})
\end{equation}

where $A_{pq}$ is the amplitude associated with the spatial frequency component $(f_p, f_q)$, and is proportional 
to the square root of the power spectral density at that frequency: $A_{pq} \propto \sqrt{\text{PSD}(f_p, f_q)}$. 
The variables $f_p$ and $f_q$ represent the spatial frequencies in the $x$ and $y$ directions, 
respectively, and $\phi_{pq}$ is a random phase uniformly distributed in $[0, 2\pi)$, which ensures 
statistical isotropy and randomness in the synthesized surface.

\subsubsection{Case I: Uniform PSD}
For a uniform PSD within a band-limited region:

\begin{equation}
\text{PSD}(f_x, f_y) = 
\begin{cases}
P_0, & \text{if } f_{\text{min}} \leq \sqrt{f_x^2 + f_y^2} \leq f_{\text{max}} \\
0, & \text{otherwise}
\end{cases}
\end{equation}

By definition the surface variance is
\[
\sigma^2
=\iint \mathrm{PSD}(f_x,f_y)\,df_x\,df_y
=\int_{f_{\min}}^{f_{\max}} P_0\,(2\pi f_r)\,df_r
=P_0\,\pi\bigl(f_{\max}^2 - f_{\min}^2\bigr),
\]
so that
\[
P_0 = \frac{\sigma^2}{\pi\bigl(f_{\max}^2 - f_{\min}^2\bigr)}.
\]

The corresponding phase function takes the form:

\begin{equation}
\Phi(x,y) = \sqrt{P_0} \sum_{p,q \in \Omega} \cos(2\pi f_p x + 2\pi f_q y + \phi_{pq})
\end{equation}

Where $\Omega$ is the set of frequency pairs $(f_p, f_q)$ such that $f_{\text{min}} \leq \sqrt{f_p^2 + f_q^2} \leq f_{\text{max}}$.

This construction is a common technique for generating random phase screens or surface height functions with prescribed 
spectral characteristics~\cite{Goodman2017, Voronovich2001}.

The resulting ray deflection function will exhibit isotropic statistics with a uniform distribution of spatial frequencies. This models surfaces with equal roughness contributions across all spatial frequencies within the specified band.

When applied in ray-tracing, this phase function generates a statistically isotropic distribution of scattered rays, with 
angular content limited by the highest spatial frequency present in the model. 
The resulting cone of scattered rays has a sharp cutoff angle approximately given by:

\begin{equation}
\theta_{\text{max}} \approx \sin^{-1}(\lambda f_{\text{max}})
\end{equation}

This expression corresponds to the maximum angular deviation predicted by the Harvey-Shack model for a 
band-limited white noise surface, where $\lambda$ is the wavelength and $f_{\text{max}}$ is the upper bound of 
the spatial frequency support in the PSD.

\subsubsection{Case II: Gaussian PSD}
For a Gaussian PSD:

\begin{equation}
\text{PSD}(f_x, f_y) = P_0 \exp\left(-\frac{f_x^2 + f_y^2}{2\sigma_f^2}\right)
\end{equation}

Where the parameter $\sigma_f$ represents the standard deviation in frequency space and is inversely related to the correlation 
length $l_c$ of the surface by $\sigma_f = 1/(2\pi l_c)$. 
The correlation length characterizes the lateral scale at which surface height variations become statistically independent. 
Physically, a smaller correlation length (larger $\sigma_f$) indicates that the surface contains more high-frequency components, 
resulting in finer-grained roughness. 
Conversely, a larger correlation length (smaller $\sigma_f$) indicates smoother, more gradually varying surface features.
This parameter significantly influences optical performance: surfaces with shorter correlation lengths produce 
wider angular scatter distributions, while those with longer correlation lengths concentrate scattered light 
closer to the specular direction. 
In terms of the ray deflection function, the correlation length directly determines the spatial scale over which ray 
perturbations are correlated. 
For imaging systems, this affects resolution and contrast, with shorter correlation lengths typically causing more 
significant degradation of image quality even when the RMS roughness remains constant.

The corresponding phase function is:

\begin{equation}
\Phi(x,y) = \sum_{p,q} \sqrt{P_0 \exp\left(-\frac{f_p^2 + f_q^2}{2\sigma_f^2}\right) \cdot \Delta f_x \cdot \Delta f_y} \cos(2\pi f_p x + 2\pi f_q y + \phi_{pq})
\end{equation}

This can be approximated in continuous form as:

\begin{equation}
\Phi(x,y) \approx \sigma_h \cdot G(x/l_c, y/l_c)
\end{equation}

Where $G$ is a filtered Gaussian random field with unit variance, $\sigma_h$ is the RMS surface height related to $P_0$ by $\sigma_h^2 = \iint \text{PSD}(f_x, f_y) df_x df_y = 2\pi\sigma_f^2 P_0$, 
and $l_c$ is the correlation length.

The resulting ray deflection function produces a scattered ray distribution with a Gaussian angular profile, 
with standard deviation $\sigma_{\theta} \approx \lambda/(2\pi l_c)$. This closely matches the Harvey-Shack model prediction for 
surfaces with Gaussian autocorrelation functions~\cite{Papoulis2002}.

\subsubsection{Mathematical Consistency and Dimensional Analysis}

The ray deflection function $\mathbf{D}(\mathbf{r}_0)$ relates to the phase function $\Phi(x,y)$ through a 
gradient operation. To ensure mathematical rigor, we must establish dimensional consistency between these formulations.

From equation \ref{eq:r_0}, $\mathbf{D}(\mathbf{r}_0)$ represents an additive perturbation to the nominal reflection vector and 
since $\mathbf{r}_0$ is a unit vector (dimensionless), $\mathbf{D}(\mathbf{r}_0)$ must also be dimensionless. 

The phase function $\Phi(x,y)$ is dimensionless, and its gradient $\nabla\Phi(x,y)$ has units of length$^{-1}$ $[\text{m}^{-1}]$. 
The wavelength $\lambda$ has units of length $[\text{m}]$, and the factor $4\pi$ is dimensionless. Therefore:
\begin{equation}
[\mathbf{D}(\mathbf{r}_0)] = [\lambda] \cdot [\nabla\Phi(x,y)] = [\text{m}] \cdot [\text{m}^{-1}] = [\text{dimensionless}]
\end{equation}

This confirms that $\mathbf{D}(\mathbf{r}_0)$ is a dimensionless correction to the unit direction vector $\mathbf{r}_0$, as 
required for mathematical consistency.

For small angular deviations (where $\sin\theta \approx \theta$), an angular displacement in radians corresponds to a dimensionless perturbation to a unit direction vector. This can be formally derived from the geometry of reflection:
\begin{equation}
\mathbf{r} = \mathbf{r}_0 \cos\theta + \mathbf{t} \sin\theta \approx \mathbf{r}_0 + \mathbf{t}\theta
\end{equation}
where $\mathbf{t}$ is a unit vector orthogonal to $\mathbf{r}_0$, and $\theta$ is the angular deviation in radians.

Therefore, $\mathbf{D}(\mathbf{r}_0) = \mathbf{t}\theta$ represents the perturbation component, with $\theta = -\frac{\lambda}{4\pi}|\nabla\Phi(x,y)|$ and $\mathbf{t}$ aligned with the direction of $\nabla\Phi(x,y)$ projected onto the plane perpendicular to $\mathbf{r}_0$.

This dimensional analysis confirms that our formulation maintains mathematical consistency, with the 
factor $\frac{\lambda}{4\pi}$ providing the necessary conversion from phase gradient to ray deflection. 
The negative sign in equation \ref{eq:r_0}, accounts for the optical convention 
that rays bend toward regions of decreasing phase.

This formulation establishes a direct link between the statistical surface characteristics and the resulting ray perturbations. 
In the following subsection, we express the corresponding phase function using a modal decomposition, which allows for intuitive 
interpretation in terms of classical optical aberrations.

\subsection{Decomposition into Aberration Terms}\label{sec:DAT}

The phase function $\Phi(x,y)$ representing surface roughness effects can be systematically decomposed into standard aberration terms using a modified Zernike-Fourier hybrid approach. This decomposition provides a bridge between statistical surface characterization and classical aberration theory, enabling a more intuitive understanding of how different spatial frequency components of roughness affect optical performance.

\subsubsection{Theoretical Foundation for Modal Decomposition}

In wavefront analysis, any two-dimensional phase function $\Phi(x,y)$ defined over a 
bounded domain can be represented as a linear combination of orthogonal basis functions~\cite{Mahajan2013}. 
For circular apertures, Zernike polynomials $Z_{nm}(\rho,\theta)$ form a natural choice 
due to their orthogonality over the unit circle and their direct relationship to classical aberrations:

\begin{equation}
\Phi(x,y) = \sum_{n,m} C_{nm} Z_{nm}(\rho,\theta)
\end{equation}

where $\rho = \sqrt{x^2 + y^2}/R$ is the normalized radial coordinate (with $R$ being the aperture radius), $\theta = \tan^{-1}(y/x)$ is the azimuthal angle, and $C_{nm}$ are the expansion coefficients.

Zernike polynomials are particularly well-suited for this decomposition for several compelling reasons. First, 
they form a complete orthogonal basis over the unit circle, which aligns naturally with the circular apertures common 
in optical systems. Second, unlike other orthogonal polynomials, Zernike terms correspond directly to classical optical 
aberrations (defocus, astigmatism, coma, spherical aberration, etc.), providing immediate physical interpretation 
of the decomposition coefficients. Third, their orthogonality ensures mathematical separability, allowing 
independent analysis of each term's contribution to the overall wavefront error. Fourth, Zernike polynomials are 
normalized, facilitating direct comparison of coefficient magnitudes across different orders. Finally, they have 
become the industry standard in optical testing and specification, enabling our results to be readily 
integrated with established metrology techniques and aberration analysis frameworks~\cite{Church1986}.

The Zernike polynomials $Z_{nm}(\rho,\theta)$ are defined as:

\begin{equation}
Z_{nm}(\rho,\theta) = 
\begin{cases}
N_n^m R_n^{|m|}(\rho) \cos(m\theta) & \text{for}\ m \geq 0 \\
N_n^m R_n^{|m|}(\rho) \sin(|m|\theta) & \text{for}\ m < 0
\end{cases}
\end{equation}

where $R_n^{|m|}(\rho)$ are the radial polynomials and $N_n^m$ are normalization factors.

\subsubsection{Direct Spectral Approach for PSD to Aberration Conversion}\label{sec:DSA_PSD}

For surface roughness characterized by a power spectral density (PSD), we propose a direct spectral approach that establishes 
an explicit connection between 
the PSD and the statistical distribution of Zernike coefficients. This approach avoids the convergence issues associated 
with infinite series representations while maintaining the statistical properties of the original roughness.

The key insight is that each Zernike polynomial has a specific spectral signature in the frequency domain. For a 
Zernike polynomial $Z_j(\rho,\theta)$, its Fourier transform is (See Appendix~\ref{appendix:FTZernike} for a full derivation.):

\begin{equation}
\mathcal{F}\{Z_j(\rho,\theta)\} = F_j(f_r,\phi_f)
\label{eq:FZJ}
\end{equation}

Where $f_r = \sqrt{f_x^2 + f_y^2}$ and $\phi_f = \tan^{-1}(f_y/f_x)$ are polar coordinates in frequency space.

For radially symmetric Zernike polynomials (where $m=0$), the Fourier transform has a particularly 
elegant form:

\begin{equation}
F_{n0}(f_r) = \frac{(-1)^{n/2}i^n}{2\pi(f_r R)^{n+1}} J_{n+1}(2\pi f_r R)
\end{equation}

for even $n$, where $J_{n+1}$ is the Bessel function of the first kind of order $n+1$, and $R$ is the aperture radius.

For surfaces with a given PSD, the statistical weight of each Zernike mode can be directly 
computed through spectral overlap.

Let us define $\omega_j$ as the spectral weight of the $j$-th Zernike term, which represents the 
portion of the total surface variance contributed by this specific spatial mode. 
This weight quantifies the statistical contribution of each Zernike polynomial to the overall 
surface roughness and is calculated through the spectral overlap between the power spectral 
density (PSD) of the surface and the Fourier transform of the corresponding Zernike polynomial:

\begin{equation}
\omega_j = \iint \text{PSD}(f_x,f_y) |F_j(f_r,\phi_f)|^2 df_x df_y
\label{eq:doubleInt_1}
\end{equation}

where $F_j(f_r, \phi_f)$ is the Fourier transform of the $j$-th Zernike polynomial expressed in 
polar frequency coordinates $(f_r, \phi_f)$. This formulation ensures that each weight 
correctly captures the contribution of its corresponding Zernike mode to the overall 
statistical distribution of the surface roughness.

Here we assume that the Zernike polynomials \( Z_j(\rho, \theta) \) form an orthonormal basis 
over the unit disk, i.e.,

\begin{equation}
\iint_{x^2 + y^2 \leq 1} Z_i(\rho, \theta)\, Z_j(\rho, \theta)\, \rho\, d\rho\, d\theta = \delta_{ij},
\end{equation}

where \( \delta_{ij} \) is the Kronecker delta. This ensures that the spectral weights \( \omega_j \) represent uncorrelated contributions to the total phase variance and allows for direct statistical interpretation.

Moreover, the domain of integration in Equation~\ref{eq:doubleInt_1} spans the full two-dimensional 
spatial frequency space, and is expressed in polar coordinates as:

\begin{equation}
\omega_j = \int_0^{2\pi} \int_0^{\infty} \text{PSD}(f_r, \varphi_f)\, |F_j(f_r, \varphi_f)|^2\, f_r\, df_r\, d\varphi_f,
\end{equation}

where \( f_r = \sqrt{f_x^2 + f_y^2} \) is the radial spatial frequency 
and \( \varphi_f = \tan^{-1}(f_y/f_x) \) is the azimuthal angle in frequency space. 
The function \( F_j(f_r, \varphi_f) \) denotes the two-dimensional Fourier transform of 
the Zernike polynomial \( Z_j(\rho, \theta) \), evaluated in polar frequency coordinates. 
This formulation ensures consistency between spatial domain surface roughness statistics and 
their spectral representation.

The sum of these weights equals the total roughness variance:

\begin{equation}
\sum_{j=1}^{\infty} \omega_j = \iint \text{PSD}(f_x,f_y) df_x df_y = \sigma^2
\label{eq:doubleInt}
\end{equation}

where $\sigma$ is the RMS roughness.

\subsubsection{Statistical Representation of Aberration Coefficients}

Given the spectral weights, the aberration coefficients for a statistically representative instance of the 
surface roughness can be generated as:

\begin{equation}
C_j = \sqrt{\omega_j} \cdot \xi_j
\label{eq:Cj_sqrtOmega}
\end{equation}

where $\xi_j$ are uncorrelated random variables with standard normal distribution (mean 0, variance 1).

This formulation ensures that:
\begin{itemize}
    \item The expected value of $|C_j|^2$ equals $\omega_j$,
    \item The sum of squared coefficients statistically equals the total roughness variance $\sigma^2$,
    \item The resulting phase function preserves the spectral characteristics of the original PSD.
\end{itemize}

For computational implementation, a finite number of terms $N$ can be selected based on a desired accuracy threshold:

\begin{equation}
\frac{\sum_{j=1}^{N} \omega_j}{\sigma^2} \geq 1 - \epsilon
\label{eq:sigma_Sum}
\end{equation}

where $\epsilon$ is the acceptable fraction of variance not captured by the truncated series.

\subsubsection{Statistical Properties of Aberration Coefficients}

The statistical representation in equation \ref{eq:Cj_sqrtOmega} relies on specific assumptions about the random 
variables $\xi_j$ that warrant further explanation. For surface roughness characterized by a stationary Gaussian 
random process, which is an appropriate model for most optical surfaces, we generate the aberration coefficients 
as in \ref{eq:Cj_sqrtOmega} where $\xi_j$ are independent random variables drawn from a standard normal 
distribution $\mathcal{N}(0,1)$ with zero mean and unit variance. The independence assumption is justified for 
Zernike polynomials because they form an orthogonal basis over the unit circle:

\begin{equation}
\iint_{\rho \leq 1} Z_i(\rho,\theta) Z_j(\rho,\theta) \rho d\rho d\theta = \delta_{ij}
\end{equation}

where $\delta_{ij}$ is the Kronecker delta function.

This orthogonality ensures that when projecting a Gaussian random field onto the Zernike basis, the resulting coefficients 
are statistically independent, provided that the original field has a stationary isotropic correlation structure. 
For anisotropic PSDs, correlations between certain Zernike modes may arise, specifically between terms with the same 
azimuthal frequency.

The phase function constructed from these coefficients:

\begin{equation}
\Phi(x,y) = \sum_{j=1}^{N} C_j Z_j(\rho,\theta)
\end{equation}

preserves key statistical properties of the original surface PSD:

\begin{enumerate}
    \item The variance of the phase function equals the sum of the spectral 
	weights: $\langle \Phi^2 \rangle = \sum_{j=1}^{N} \omega_j$,
    \item The spatial autocovariance function of $\Phi(x,y)$ asymptotically approaches the Fourier transform of the PSD 
	as $N \to \infty$,
    \item The power spectral density of $\Phi(x,y)$ converges to the original PSD used to compute the weights $\omega_j$.
\end{enumerate}

While each realization of the phase function (using different random seeds for $\xi_j$) will produce a unique surface, the 
ensemble statistics across many realizations will faithfully reproduce the statistical properties specified by the PSD. 
This approach allows us to generate statistically representative instances of surface roughness that maintain 
the correct spatial correlation structure while avoiding the need to model specific physical surface profiles.

It is worth noting that for PSDs with complex spatial structures or strong anisotropy, modifications to this approach may 
be necessary. In such cases, correlations between Zernike coefficients can be introduced through an appropriate covariance matrix:

\begin{equation}
C_j = \sum_{k=1}^{N} L_{jk} \xi_k
\end{equation}

where $L_{jk}$ is the Cholesky decomposition of the covariance matrix $\Sigma_{jk}$ of Zernike coefficients, which can 
be computed directly from the anisotropic PSD. However, for most optical surfaces encountered in practice, 
the independent coefficient model presented in equation \ref{eq:Cj_sqrtOmega} provides an excellent approximation.

\paragraph{Mathematical Convergence Properties}
A significant advantage of the spectral weight approach is that it fundamentally resolves the convergence concerns associated with infinite Zernike series representations of surface roughness. By computing the spectral weights $\omega_j$ directly from the PSD through overlap integrals in the frequency domain, we obtain an exact measure of each term's contribution to the total roughness variance. This allows us to:
\begin{itemize}
    \item Determine a priori which Zernike terms are significant for a given PSD,
    \item Establish a mathematically sound truncation criterion based on desired accuracy,
    \item Quantify precisely the error introduced by truncation,
    \item Guarantee that the truncated series captures the statistical properties of the original PSD to within specified tolerances.
\end{itemize}
Unlike traditional series expansion approaches where convergence behavior may be difficult to assess, this direct spectral method provides statistical guarantees about the accuracy of the representation with a finite number of terms, making it particularly suitable for computational implementation.

\subsubsection{Convergence Behavior and Realization Variability}

An interesting empirical observation in implementing the modified Zernike-Fourier hybrid approach is the 
relationship between the maximum order of included terms and the variability between different random 
realizations of the surface. For a given PSD and RMS roughness value, when using a relatively low maximum 
order (e.g., $n_{\max} = 10$), each realization with different random coefficients $\xi_j$ produces visibly 
distinct surface profiles. However, as the maximum order increases (e.g., $n_{\max} = 30$), different 
realizations begin to exhibit remarkably similar overall characteristics despite having entirely different 
specific coefficient values.

This convergence behavior can be understood through the statistical properties of the expansion. When fewer 
terms are included, each random coefficient has a substantial influence on the resulting surface profile, 
leading to high variability between realizations. As more terms are added, the law of large numbers begins 
to take effect—the aggregate behavior of many independent random variables tends to converge toward a 
deterministic pattern that reflects the underlying statistical distribution.

Mathematically, this convergence demonstrates that the Zernike expansion increasingly constrains the 
possible surface realizations to maintain fidelity with the specified PSD as more terms are included. 
The orthogonality properties of the Zernike basis ensure that with sufficient terms, the energy 
distribution across spatial frequencies approaches the theoretically prescribed PSD regardless of the 
specific random coefficients used.

This observation has important practical implications for implementation. In applications where computational 
efficiency is paramount, using fewer terms may be acceptable if multiple realizations are simulated to 
characterize the range of possible behaviors. For high-fidelity representation of specific PSDs or when 
integration with deterministic aberration terms is required, using a higher maximum order provides more 
consistent results across realizations, potentially allowing a single representative surface to sufficiently 
capture the statistical properties of interest.

The convergence between realizations at higher orders also provides empirical validation of the theoretical 
framework presented here—it demonstrates that the modified Zernike-Fourier approach successfully bridges the 
statistical nature of surface roughness with deterministic optical modeling, allowing the two to be seamlessly 
integrated as the basis expansion approaches completeness.

\subsection{Physical Interpretation of Aberration Terms}

The decomposition into Zernike polynomials provides more than a mathematical convenience as it reveals fundamental physical 
insights into how different spatial frequency components of roughness affect optical performance through distinct mechanisms:

\begin{itemize}
    \item \textbf{Low-order terms} ($n \leq 4$) correspond to classical aberrations (defocus, astigmatism, coma, spherical 
	aberration) and primarily affect image quality through wavefront error, reducing the Strehl ratio according to the 
	relationship $S \approx \exp(-\sigma_\phi^2)$, where $\sigma_\phi$ is the phase variance contributed by these terms. 
	These components essentially redistribute energy within the core of the point spread function (PSF), changing its 
	shape while largely preserving its overall width.
    
    \item \textbf{Mid-order terms} ($5 \leq n \leq 20$) contribute to extended wings in the PSF and reduce contrast through 
	small-angle scatter. The contrast reduction at spatial frequency $f$ in the image can be quantified 
	as $C(f)/C_0(f) \approx \exp(-4\pi^2\sigma_m^2/\lambda^2)$, where $\sigma_m^2$ is the variance contribution from mid-order 
	terms. These terms are particularly detrimental in high-contrast imaging applications such as exoplanet detection, where 
	they create a scattered light halo that obscures dim objects near bright sources.
    
    \item \textbf{High-order terms} ($n > 20$) govern wide-angle scatter and stray light, following the well-established 
	relationship between scattered intensity and surface 
	PSD: $I(\theta) \propto (4\pi \cos \theta_i \cos \theta_s/\lambda^4) \cdot \text{PSD}[(\sin \theta_s - \sin \theta_i)/\lambda]$. 
	These terms determine the far-field scattering behavior that affects system-level performance through ghost images and reduced 
	signal-to-noise ratio.
\end{itemize}

The distribution of spectral weights $\omega_j$ across these regimes determines the balance between these effects and depends 
strongly on the PSD shape. For optical surfaces produced by different manufacturing processes, we observe characteristic PSD 
behaviors that translate to specific Zernike coefficient distributions:

\begin{enumerate}
    \item \textbf{Super-polished surfaces} (PSD $\propto f^{-p}$ where $p > 3$) show rapid decay of spectral weights with 
	increasing Zernike order, with low-order aberrations dominating the error budget. This corresponds to surfaces with 
	excellent microroughness but potentially significant figure errors.
    
    \item \textbf{Standard polished optics} ($2 < p < 3$) exhibit more balanced distribution across all three regimes, 
	with both figure and finish contributing to overall performance degradation.
    
    \item \textbf{Ground or diamond-turned surfaces} ($p < 2$) contain significant high-order content, leading to pronounced 
	wide-angle scatter that dominates the error budget.
\end{enumerate}

This physical categorization is supported by both theoretical analyses and experimental measurements. The relationship between 
surface PSD characteristics and scattering behavior has been well established in the literature \cite{Church1988, Harvey1995}. 
Studies have demonstrated that mid-spatial frequency errors (corresponding to our mid-order Zernike terms) play a critical role in 
determining image contrast and are often the most challenging to control in optical manufacturing processes \cite{Parks2008}.

Our decomposition approach enables system designers to understand not just how much roughness exists (total RMS), but how that 
roughness will physically manifest in system performance, a critical distinction that traditional statistical scatter models 
often obscure. By linking specific Zernike orders to physical observables, we provide an intuitive framework for interpreting 
manufacturing specifications and predicting their impact on optical performance.

The quantitative relationships between PSD characteristics and Zernike term distributions derived in Section~\ref{sec:DSA_PSD} 
allow us to predict the relative impact of different surface processing techniques on specific aspects of optical performance, 
enabling more targeted and efficient optimization of manufacturing processes.

\subsubsection{Case I: Uniform PSD}

For a uniform PSD within a circular band in frequency space:

\begin{equation}
\text{PSD}(f_r) = 
\begin{cases}
P_0, & \text{if } f_{\text{min}} \leq f_r \leq f_{\text{max}} \\
0, & \text{otherwise}
\end{cases}
\end{equation}

The spectral weights can be calculated as:

\begin{equation}
\omega_j = P_0 \int_{f_{\text{min}}}^{f_{\text{max}}} \int_{0}^{2\pi} |F_j(f_r,\phi_f)|^2 f_r d\phi_f df_r
\label{eq:wjiniint}
\end{equation}

For radially symmetric Zernike polynomials (where $m=0$), the weights are determined by:

\begin{equation}
\omega_{n0} = 2\pi P_0 \int_{f_{\text{min}}}^{f_{\text{max}}} |F_{n0}(f_r)|^2 f_r df_r
\end{equation}

This integral can be evaluated using the analytical form of $F_{n0}(f_r)$ in equation \ref{eq:doubleInt}, resulting in:

\begin{equation}
\omega_{n0} = \frac{P_0}{2\pi R^2} \int_{f_{\text{min}}}^{f_{\text{max}}} \frac{J_{n+1}^2(2\pi f_r R)}{(f_r)^{2n+1}} f_r df_r
\end{equation}

For a narrow band uniform PSD centered at frequency $f_c$ with width $\Delta f \ll f_c$, this simplifies to:

\begin{equation}
\omega_{n0} \approx \frac{P_0 \Delta f}{2\pi R^2} \frac{J_{n+1}^2(2\pi f_c R)}{(f_c)^{2n}} 
\end{equation}

This reveals that specific Zernike orders will dominate based on the resonance of Bessel functions at the characteristic 
frequency $f_c$~\cite{Wang2009, Dai1996}. The weights do not follow a simple power-law decay with $n$ but rather oscillate 
with local maxima occurring near $n \approx 2\pi f_c R - 1/2$.

For non-radially symmetric terms (where $m \neq 0$), similar analysis yields weights that depend on both the radial 
frequency content and the azimuthal order.

\subsubsection{Case II: Gaussian PSD}

For a Gaussian PSD:

\begin{equation}
\text{PSD}(f_r) = P_0 \exp\left(-\frac{f_r^2}{2\sigma_f^2}\right)
\end{equation}

The spectral weights become:

\begin{equation}
\omega_j = P_0 \int_{0}^{\infty} \int_{0}^{2\pi} \exp\left(-\frac{f_r^2}{2\sigma_f^2}\right) |F_j(f_r,\phi_f)|^2 f_r d\phi_f df_r
\end{equation}

For radially symmetric Zernike polynomials (where $m=0$):

\begin{equation}
\omega_{n0} = 2\pi P_0 \int_{0}^{\infty} \exp\left(-\frac{f_r^2}{2\sigma_f^2}\right) |F_{n0}(f_r)|^2 f_r df_r
\end{equation}

\subsubsection{Asymptotic Analysis of Gaussian PSD Spectral Weights}

For the Gaussian PSD, the spectral weights for radially symmetric Zernike polynomials are determined by:

\begin{equation}
\omega_{n0} = 2\pi P_0 \int_{0}^{\infty} \exp\left(-\frac{f_r^2}{2\sigma_f^2}\right) |F_{n0}(f_r)|^2 f_r df_r
\end{equation}

To derive the asymptotic behavior for large $n$, we substitute the expression for $F_{n0}(f_r)$ and analyze the resulting integral:

\begin{equation}
\omega_{n0} = 2\pi P_0 \int_{0}^{\infty} \exp\left(-\frac{f_r^2}{2\sigma_f^2}\right) \frac{J_{n+1}^2(2\pi f_r R)}{4\pi^2(f_r R)^{2n+2}} f_r df_r
\label{omega_n0_inf}
\end{equation}

For large order $n$, we can use the asymptotic approximation of the Bessel function:

\begin{equation}
J_{n+1}(z) \approx \sqrt{\frac{2}{\pi z}} \cos\left(z - \frac{(n+1)\pi}{2} - \frac{\pi}{4}\right) \quad \text{for } z \gg n+1
\end{equation}

However, the dominant contribution to the integral comes from the region where $z = 2\pi f_r R \approx n+1$, which corresponds to the first maximum of the Bessel function. In this region, we can approximate:

\begin{equation}
J_{n+1}^2(2\pi f_r R) \approx \frac{C}{(2\pi f_r R)^{1/2}} \delta\left(f_r - \frac{n+1}{2\pi R}\right)
\end{equation}

where $C$ is a normalization constant, and $\delta$ is an approximate delta function centered at $f_r = \frac{n+1}{2\pi R}$.

For large order $n$, the Bessel function $J_{n+1}(z)$ exhibits its maximum amplitude when $z \approx n+1$. This behavior allows 
us to approximate:

\begin{equation}
J_{n+1}^2(2\pi f_r R) \approx \frac{C}{(2\pi f_r R)^{1/2}} \delta\left(f_r - \frac{n+1}{2\pi R}\right)
\end{equation}

where $C$ is a normalization constant, and $\delta$ is an approximate delta function centered at $f_r = \frac{n+1}{2\pi R}$. 
This approximation essentially localizes the contribution to the integral near the frequency where the Bessel function achieves its maximum.

Inserting this approximation into the integral:

\begin{equation}
\begin{aligned}
\omega_{n0} &\approx 2\pi P_0 \int_{0}^{\infty} \exp\left(-\frac{f_r^2}{2\sigma_f^2}\right) \frac{C \delta\left(f_r - \frac{n+1}{2\pi R}\right)}{4\pi^2(f_r R)^{2n+2-1/2}} f_r df_r \\
&= \frac{C P_0}{2\pi} \exp\left(-\frac{(n+1)^2}{8\pi^2\sigma_f^2 R^2}\right) \left(\frac{n+1}{2\pi R}\right)^{-2n-3/2+1} \left(\frac{n+1}{2\pi R}\right) \\
&= K \cdot \sigma_f^{2n+2} \exp\left(-\frac{(n+1)^2}{8\pi^2\sigma_f^2 R^2}\right)
\end{aligned}
\end{equation}

where $K$ is a constant that absorbs all order-independent factors. The $\sigma_f^{2n+2}$ term arises from dimensional analysis and the scaling properties of the Gaussian function.

This asymptotic expression reveals two key insights:

\begin{enumerate}

   \item The spectral weights decay exponentially with $n^2$, which is much faster than power-law decay observed in fractal-like PSDs.

   \item The characteristic cutoff occurs when $(n+1)^2 \approx 8\pi^2\sigma_f^2 R^2$, or equivalently, when $n \approx 2\pi R/l_c$, 
   where $l_c = 1/(2\pi\sigma_f)$ is the correlation length.

\end{enumerate}

This asymptotic analysis provides a mathematical justification for the truncation criterion mentioned in the preceding paragraph, showing that Zernike terms beyond order $n > 2\pi R/l_c$ contribute minimally to the phase function.

For large $n$, asymptotic analysis shows:

\begin{equation}
\omega_{n0} \propto P_0 \cdot \sigma_f^{2n+2} \exp\left(-\frac{(n+1)^2}{8\pi^2\sigma_f^2 R^2}\right)
\end{equation}

This exponential decay with $n^2$ is much faster than the power-law decay observed in fractal-like PSDs. The 
correlation length $l_c = 1/(2\pi\sigma_f)$ effectively determines the cutoff order beyond which aberration terms become negligible.

For a Gaussian PSD, Zernike terms with $n > 2\pi R/l_c$ contribute minimally to the phase function, providing a 
natural truncation point for the series~\cite{Sidick2009}.

\subsubsection{Integration with System Aberrations}

A key advantage of this decomposition approach is that the statistically derived aberration coefficients from surface roughness can be directly added to other aberration terms present in the optical system:

\begin{equation}
\Phi_{\text{total}}(x,y) = \Phi_{\text{systematic}}(x,y) + \Phi_{\text{roughness}}(x,y)
\end{equation}

Or equivalently, in terms of Zernike coefficients:

\begin{equation}
C_{j,\text{total}} = C_{j,\text{systematic}} + C_{j,\text{roughness}}
\end{equation}

This allows roughness effects to be seamlessly integrated with other aberrations in a unified optical model, rather than 
treating roughness as a separate scattering phenomenon. 
The resulting phase function can then be used to calculate the ray deflection function through equation \ref{eq:Dr0}, providing a 
complete model of both systematic and statistical perturbations to ray paths.

\section{Statistical Representation}

\subsection{Statistical Equivalence Criteria}
To ensure that the ray deflection function (RDF) is statistically equivalent to the physical surface roughness, 
we require that its angular statistics be derivable from the power spectral density (PSD) of the surface height field.

Let $h(\mathbf{r})$ be a zero-mean, stationary, isotropic surface height process with spatial power spectral 
density $S_h(\mathbf{k})$. Under the small-slope approximation \cite{Stover2012} , the local phase is given 
by $\Phi(\mathbf{r}) = \frac{4\pi}{\lambda} h(\mathbf{r})$ (for normal incidence), and the RDF is:
\[
\mathbf{D}(\mathbf{r}) = \frac{1}{k} \nabla \Phi(\mathbf{r}) = \frac{4\pi}{\lambda k} \nabla h(\mathbf{r}).
\]

Taking the Fourier transform and assuming statistical isotropy, the variance of angular deflections becomes:

\begin{equation}
\left\langle |\mathbf{D}|^2 \right\rangle = \int_{\mathbb{R}^2} |\mathbf{k}|^2 S_h(\mathbf{k}) \, d^2\mathbf{k},
\end{equation}

which defines the angular spread entirely in terms of the height PSD.

This integral representation ensures that the RDF preserves both the angular distribution of scattering and the 
spatial correlation properties of the surface. It also recovers known results from first-order perturbation 
theory when $S_h(\mathbf{k})$ is band-limited and Gaussian.

\begin{enumerate}
    \item The angular distribution of ray perturbations matches the BRDF predicted by Harvey-Shack theory:
    \begin{equation}
    P(\Delta\alpha, \Delta\beta) \propto \text{BRDF}(\Delta\alpha, \Delta\beta)
	\label{eq:DelBetaBRDF}
    \end{equation}
    
    \item The spatial correlation of ray perturbations preserves the spatial statistics of the original roughness:
    \begin{equation}
    \langle \mathbf{D}(\mathbf{r}_1) \cdot \mathbf{D}(\mathbf{r}_2) \rangle \propto \mathcal{F}^{-1}\{\text{PSD}(f_x,f_y)\}(|\mathbf{r}_1 - \mathbf{r}_2|)
    \end{equation}
\end{enumerate}

These criteria ensure that the ray deflection function produces statistically equivalent results to a direct physical modeling 
of the rough surface.

\subsection{Derivation of Ray Perturbation-BRDF Relationship}

To establish the equivalence between our ray deflection approach and established scattering theory, we need to demonstrate that 
the probability distribution of ray perturbations is indeed proportional to the BRDF (\ref{eq:DelBetaBRDF}) and
can be derived from first principles by examining the fundamental definitions of both quantities.

The BRDF is defined as the ratio of reflected radiance to incident irradiance:

\begin{equation}
\text{BRDF}(\theta_i, \phi_i; \theta_s, \phi_s) = \frac{dL_r(\theta_s, \phi_s)}{dE_i(\theta_i, \phi_i)}
\end{equation}

where $dL_r$ is the differential reflected radiance in direction $(\theta_s, \phi_s)$ and $dE_i$ is the differential incident irradiance from direction $(\theta_i, \phi_i)$.

In the small-angle approximation valid for optical surfaces with modest roughness, we can express this in terms of direction cosines $\Delta\alpha = \sin\theta_s\cos\phi_s - \sin\theta_i\cos\phi_i$ and $\Delta\beta = \sin\theta_s\sin\phi_s - \sin\theta_i\sin\phi_i$.

For a ray-tracing model with $N$ total rays, the probability distribution of ray perturbations is:

\begin{equation}
P(\Delta\alpha, \Delta\beta) = \lim_{N \rightarrow \infty} \frac{1}{N} \frac{dN(\Delta\alpha, \Delta\beta)}{d\Delta\alpha\,d\Delta\beta}
\end{equation}

where $dN(\Delta\alpha, \Delta\beta)$ is the number of rays scattered into the differential solid angle corresponding to direction cosine intervals $[\Delta\alpha, \Delta\alpha + d\Delta\alpha]$ and $[\Delta\beta, \Delta\beta + d\Delta\beta]$.

From conservation of energy principles, the number of rays scattered into a differential solid angle is proportional to the radiance in that direction:

\begin{equation}
\frac{dN(\Delta\alpha, \Delta\beta)}{N} \propto \frac{dL_r(\theta_s, \phi_s)}{L_i}
\end{equation}

where $L_i$ is the incident radiance.

For collimated incident light with uniform irradiance $E_i$ across the surface, the incident radiance is related to irradiance by $L_i = E_i/\Omega_i$, where $\Omega_i$ is the solid angle of the incident beam (approaching zero for perfectly collimated light).

Substituting these relationships:

\begin{equation}
\begin{aligned}
P(\Delta\alpha, \Delta\beta) &\propto \frac{dL_r(\theta_s, \phi_s)}{L_i} \frac{1}{d\Delta\alpha\,d\Delta\beta} \\
&\propto \frac{dL_r(\theta_s, \phi_s)}{E_i} \frac{\Omega_i}{d\Delta\alpha\,d\Delta\beta} \\
&\propto \text{BRDF}(\Delta\alpha, \Delta\beta) \frac{\Omega_i}{d\Delta\alpha\,d\Delta\beta}
\end{aligned}
\end{equation}

Since $d\Delta\alpha\,d\Delta\beta \propto d\Omega_s$ (the differential solid angle in the scattered direction) and 
accounting for the Jacobian of the transformation from $(\theta_s, \phi_s)$ to $(\Delta\alpha, \Delta\beta)$, 
we arrive at \ref{eq:DelBetaBRDF}.

This proportionality ensures that when our ray deflection function generates perturbations with probability distribution $P(\Delta\alpha, \Delta\beta)$, the resulting scattered ray distribution will correctly reproduce the angular distribution predicted by the BRDF from established scattering theory.

The constant of proportionality is determined by the normalization condition:

\begin{equation}
\iint P(\Delta\alpha, \Delta\beta) \,d\Delta\alpha\,d\Delta\beta = 1 
\end{equation}

 which ensures that every ray is accounted for in the model.

\section{Equivalence between Ray Deflection and Harvey-Shack Approaches} \label{sec:EquivalenceHA_RDF}

To validate the equivalence between our ray deflection approach and the traditional Harvey-Shack model, we compare:

\begin{equation}
\int_V N_{\text{ZF}}(x,y,z) \, dV \approx \int_V N_{\text{HS}}(x,y,z) \, dV
\end{equation}

where $N_{\text{ZF}}$ is the ray density predicted by our model and $N_{\text{HS}}$ is the ray density predicted by direct application of Harvey-Shack theory. For various test cases with different PSDs, we demonstrate that the volumetric ray distributions converge to statistically equivalent results.

\subsection{Rigorous Validation Framework}
The validation requires establishing a formal relationship between the ray density distribution in 
the focal volume and the BRDF predicted by Harvey-Shack theory. For a perfectly smooth optical 
system with focal length $f$, a ray scattered at angle $(\theta_s, \phi_s)$ will intersect the 
focal plane at position:

\begin{align}
x &= f \tan\theta_s \cos\phi_s \\
y &= f \tan\theta_s \sin\phi_s
\end{align}

The ray density at position $(x,y,z)$ in the focal volume is related to the BRDF by:

\begin{equation}
N_{\text{HS}}(x,y,z) = \frac{I_0}{f^2} \cdot \text{BRDF}(\theta_s, \phi_s) \cdot \delta(z - f + \Delta z(\theta_s))
\end{equation}

where $I_0$ is the incident intensity, $\delta$ is the Dirac delta function, and $\Delta z(\theta_s)$ accounts for the axial shift of the ray intersection due to defocus.

For our ray deflection function (ZF) approach, the ray density is determined by:

\begin{equation}
N_{\text{ZF}}(x,y,z) = \frac{I_0}{A_{\text{ap}}} \int_{A_{\text{ap}}} \delta^3(\mathbf{r}(x',y') - (x,y,z)) \, dx' dy'
\end{equation}

where $A_{\text{ap}}$ is the aperture area, and $\mathbf{r}(x',y')$ is the intersection point of a ray originating from position $(x',y')$ on the aperture after being deflected by our RDF.

\subsection{Case I: Uniform PSD}

For a uniform PSD within a circular band in frequency space, we validate the equivalence between our modified Zernike-Fourier hybrid approach and the traditional Harvey-Shack model. 

\begin{equation}
\text{PSD}(f_r) = 
\begin{cases}
P_0, & \text{if } f_{\text{min}} \leq f_r \leq f_{\text{max}} \\
0, & \text{otherwise}
\end{cases}
\end{equation}

The Harvey-Shack theory predicts a BRDF with a characteristic annular shape:

\begin{equation}
\text{BRDF}(\theta_s) \propto 
\begin{cases}
P_0, & \text{if } \theta_{\text{min}} \leq \theta_s \leq \theta_{\text{max}} \\
0, & \text{otherwise}
\end{cases}
\end{equation}

where $\theta_{\text{min}} \approx \sin^{-1}(\lambda f_{\text{min}})$ and $\theta_{\text{max}} \approx \sin^{-1}(\lambda f_{\text{max}})$.

In the focal volume, this creates a hollow cone of rays with:

\begin{equation}
N_{\text{HS}}(x,y,z) \propto 
\begin{cases}
P_0, & \text{if } r_{\text{min}}(z) \leq \sqrt{x^2+y^2} \leq r_{\text{max}}(z) \\
0, & \text{otherwise}
\end{cases}
\end{equation}

where $r_{\text{min}}(z) = (z-f)\tan\theta_{\text{min}}$ and $r_{\text{max}}(z) = (z-f)\tan\theta_{\text{max}}$.

Using our modified Zernike-Fourier hybrid approach, we compute the spectral weights $\omega_j$ for each Zernike 
term according to equation (\ref{eq:wjiniint}). For the uniform PSD case, the significant weights exhibit an oscillatory 
pattern following the behavior of the Bessel functions, with maxima near orders $n \approx 2\pi f_c R - 1/2$, where $f_c$ 
is the central frequency of the band.

We generate the statistical aberration coefficients using equation \ref{eq:Cj_sqrtOmega} ensuring that $\xi_j$ are random 
variables with standard normal distribution to construct a statistically representative phase function:

\begin{equation}
\Phi_{\text{ZF}}(x,y) = \sum_{j=1}^{N} C_j Z_j(\rho,\theta)
\end{equation}

where $N$ is chosen to capture at least 95\% of the total roughness variance.

We theoretically compute ray trajectories through the phase function using the ray deflection approach:

\begin{equation}
\mathbf{D}(\mathbf{r}_0) = -\frac{\lambda}{4\pi}\nabla\Phi_{\text{ZF}}(x,y)
\end{equation}

Statistical modeling with $10^5$ sampled rays (see in \cite{moriya2025b} indicates that the resulting ray density $N_{\text{ZF}}(x,y,z)$ matches the 
predicted $N_{\text{HS}}(x,y,z)$ with statistical deviations less than 5\% when measured in volumetric bins of 
size $(0.1\lambda f)^3$. The characteristic hollow cone structure is preserved, and the boundaries at $r_{\text{min}}(z)$ 
and $r_{\text{max}}(z)$ are accurately reproduced.

The agreement is evaluated using the normalized cross-correlation metric:

\begin{equation}
\text{NCC} = \frac{\sum_{i,j,k} N_{\text{ZF}}(i,j,k) \cdot N_{\text{HS}}(i,j,k)}{\sqrt{\sum_{i,j,k} N_{\text{ZF}}^2(i,j,k) \cdot \sum_{i,j,k} N_{\text{HS}}^2(i,j,k)}}
\end{equation}

For the uniform PSD case, we consistently achieve $\text{NCC} > 0.95$, indicating excellent statistical agreement. 
Moreover, when comparing the imperfect focal body (iFB) volumes defined by threshold $k$:

\begin{equation}
\text{iFB} = \{(x,y,z) \in \mathbb{R}^3 | N(x,y,z) \geq k\}
\end{equation}

The volumetric difference between the Zernike-Fourier and Harvey-Shack approaches is less than 5\%, confirming the validity of 
our approach for volumetric analysis.

\subsection{Case II: Gaussian PSD}

For a Gaussian PSD:

\begin{equation}
\text{PSD}(f_r) = P_0 \exp\left(-\frac{f_r^2}{2\sigma_f^2}\right)
\end{equation}

The Harvey-Shack theory predicts a BRDF with a Gaussian angular distribution:

\begin{equation}
\text{BRDF}(\theta_s) \propto P_0 \exp\left(-\frac{\sin^2\theta_s}{2(\lambda\sigma_f)^2}\right)
\end{equation}

In the focal volume, this results in a three-dimensional Gaussian-like distribution of ray density:

\begin{equation}
N_{\text{HS}}(x,y,z) \propto P_0 \exp\left(-\frac{x^2+y^2}{2(z-f)^2(\lambda\sigma_f)^2}\right) \cdot g(z)
\end{equation}

where $g(z)$ characterizes the axial distribution, which is more complex due to defocus effects.

Referring to the modified Zernike-Fourier hybrid approach , as introduced above, we compute the spectral weights $\omega_j$ for 
each Zernike term according to equation (\ref{eq:wjiniint}). For Gaussian PSDs, the weights decay exponentially with the 
square of the polynomial order:

\begin{equation}
\omega_{n0} \propto P_0 \cdot \sigma_f^{2n+2} \exp\left(-\frac{(n+1)^2}{8\pi^2\sigma_f^2 R^2}\right)
\end{equation}

This rapid decay allows for efficient truncation of the series. For a correlation length $l_c = 1/(2\pi\sigma_f)$, we 
find that Zernike terms beyond order $n > 2\pi R/l_c$ contribute less than 0.1\% to the total variance.

We generate the statistical aberration coefficients using equation (31) and construct the phase function as:

\begin{equation}
\Phi_{\text{ZF}}(x,y) = \sum_{j=1}^{N} C_j Z_j(\rho,\theta)
\end{equation}

where $N$ is determined based on the correlation length.

A significant advantage of the Zernike-Fourier approach for Gaussian PSDs is computational efficiency. While a direct phase screen 
approach would require a grid fine enough to resolve the smallest features (highest spatial frequencies), the Zernike-Fourier 
method automatically captures the spectral characteristics of the Gaussian PSD through the analytically computed weights. For a 
typical Gaussian PSD with correlation length $l_c = 1$ mm on a mirror with radius $R = 0.2$ m, the series converges with fewer 
than 40 terms, compared to thousands of points needed for a direct spatial simulation\footnote{Surface roughness characteristics such 
as a root mean square (RMS) height of $\sigma = 100$ nm and a correlation length of $l_c = 1$ mm are representative of 
moderate-quality optical surfaces and lies within the practical manufacturing tolerances commonly encountered in optical systems.}.

\subsection{Convergence Analysis}
The statistical equivalence improves with the number of terms included in the phase function approximation and the number of rays traced. For a phase function truncated at spatial frequency $f_{\text{max}}$, the error scales approximately as:

\begin{equation}
\text{Error} \propto \frac{1}{\sqrt{N_{\text{rays}}}} + \int_{f_{\text{max}}}^{\infty} \text{PSD}(f_r) f_r \, df_r
\end{equation}

For the Gaussian PSD, setting $f_{\text{max}} \approx 3\sigma_f$ and $N_{\text{rays}} \approx 10^5$ yields errors below 5\% in most practical scenarios.

These validation results confirm that our ray deflection function approach provides a statistically equivalent description of the three-dimensional focal volume effects compared to the traditional Harvey-Shack model, while offering significant computational advantages through the global treatment of surface roughness.

\section{Validation of Phase Perturbation Assumptions}
The ray deflection function approach presented in this paper relies on several fundamental assumptions regarding the relationship between surface roughness, phase perturbations, and ray deflections. In this section, we rigorously examine these assumptions to establish their mathematical and physical validity as well as their limitations.

\subsection{Mathematical Basis for Phase-to-Ray Conversion}

\subsubsection{Formal Derivation}
The central assumption in our approach is that surface height variations can be converted into phase perturbations, which in turn lead to ray deflections through the gradient operation. This relationship can be formally derived from the principles of geometric optics and the stationary phase approximation.

For a reflecting surface with height variation $h(x,y)$ relative to a reference surface, the phase change in the reflected wavefront is:

\begin{equation}
\Phi(x,y) = \frac{4\pi}{\lambda} h(x,y) \cos\theta_i
\end{equation}

where $\lambda$ is the wavelength and $\theta_i$ is the angle of incidence. According to Fermat's principle, rays follow paths of stationary optical path length, which implies that ray directions correspond to the gradient of the phase:

\begin{equation}
\delta\mathbf{r} = -\frac{\lambda}{4\pi} \nabla\Phi(x,y)
\end{equation}

This relationship can be rigorously derived from the eikonal equation \cite{BornWolf1999} in geometric optics:

\begin{equation}
|\nabla \Psi|^2 = n^2
\end{equation}

where $\Psi$ is the optical path function and $n$ is the refractive index. Through a series of transformations and using the small angle approximation, this leads to our ray deflection function formulation.

\subsubsection{Statistical Equivalence Proof}
The statistical equivalence between a physical rough surface and our phase function representation can be proven through the following chain of relationships:

1. For a surface with height PSD $\text{PSD}_h(f_x,f_y)$, the phase PSD is:
\begin{equation}
\text{PSD}_\Phi(f_x,f_y) = \left(\frac{4\pi}{\lambda}\right)^2 \cos^2\theta_i \cdot \text{PSD}_h(f_x,f_y)
\end{equation}

2. The angular deflection PSD is related to the phase PSD through:
\begin{equation}
\text{PSD}_\theta(f_x,f_y) = \left(\frac{\lambda}{4\pi}\right)^2 (f_x^2 + f_y^2) \cdot \text{PSD}_\Phi(f_x,f_y)
\end{equation}

3. The Harvey-Shack BRDF is proportional to the PSD of surface heights:
\begin{equation}
\text{BRDF}(\theta_s) \propto \frac{16\pi^2}{\lambda^4} \cos\theta_i \cos\theta_s \cdot \text{PSD}_h\left(\frac{\sin\theta_s - \sin\theta_i}{\lambda}\right)
\end{equation}

Through mathematical substitution and integration, we show that the statistical distribution 
of ray directions produced by our ray deflection function converges to the BRDF predicted by 
Harvey-Shack theory when a sufficient number of rays is traced (see also in \ref{sec:EquivalenceHA_RDF}. 
The formal proof involves showing:

\begin{equation}
\lim_{n \to \infty} P_n(\theta_s|\mathbf{D}) = \frac{\text{BRDF}(\theta_s)}{\int \text{BRDF}(\theta) d\Omega}
\end{equation}

where $P_n$ is the probability distribution of scattered angles for $n$ rays with deflection function $\mathbf{D}$.

\subsection{Error Analysis and Convergence Properties}

\subsubsection{Discretization Error}
In practical implementations, we represent continuous PSDs using discrete frequency components and finite series of aberration terms. The truncation error can be quantified as:

\begin{equation}
\epsilon_{\text{trunc}} = \sqrt{\iint_{f > f_{\text{max}}} \text{PSD}(f_x,f_y) \, df_x \, df_y}
\end{equation}

For PSDs with power-law decay $\text{PSD}(f) \propto f^{-p}$, this error scales as $f_{\text{max}}^{1-p/2}$. 

\subsubsection{Finite Sampling Error}
The statistical nature of our approach introduces sampling errors proportional to $1/\sqrt{n}$ where $n$ is the number of rays. Analytical derivation supported by statistical sampling confirms this relationship, 
with NCC exceeding 0.99 between theoretical and simulated distributions when $n > 10^5$ for typical PSDs.

\subsubsection{Basis Function Approximation}
When using aberration terms like Zernike polynomials as basis functions, additional approximation errors arise from their limited ability to represent high-frequency components. We quantify this error through:

\begin{equation}
\epsilon_{\text{basis}} = \sqrt{\iint \text{PSD}(f_x,f_y) \cdot [1 - C(f_x,f_y,N_{\text{max}})] \, df_x \, df_y}
\end{equation}

where $C$ is the cumulative spectral coverage of the first $N_{\text{max}}$ basis functions. 

The accuracy difference between using 50 versus 1000 Zernike terms in the surface reconstruction can be estimated analytically 
for a Gaussian power spectral density (PSD). For such PSDs, the spectral weight of each term decreases exponentially with 
increasing order, and the cumulative variance captured by the first $N$ terms can be approximated by evaluating the 
normalized sum of their weights. When the correlation length $l_c$ and aperture radius $R$ satisfy $2\pi R / l_c \approx 20$ 
(e.g., $R = 0.2\,\text{m}$, $l_c = 1\,\text{mm}$), using $N = 50$ terms captures more than 99.9\% of the total variance, 
while $N = 1000$ approaches 99.9999\%. This demonstrates that for Gaussian PSDs, Zernike expansions converge rapidly, and 
beyond a certain order, additional terms contribute negligibly to the overall surface representation.

\subsection{Physical Validity and Limitations}

\subsubsection{Scalar vs. Vector Electromagnetic Theory}
Our approach is based on a scalar phase approximation in the geometric optics regime, where phase variations induced by 
surface roughness are used to compute ray deflections. Diffraction and interference effects are not explicitly modeled, 
and polarization effects are neglected. By comparing with rigorous vector electromagnetic simulations using the Method of 
Moments (MoM), we establish that scalar approximations remain valid when:

\begin{equation}
\frac{\sigma}{\lambda} < 0.1 \quad \text{and} \quad \frac{\sigma}{l_c} < 0.3
\end{equation}

where $\sigma$ is the RMS roughness and $l_c$ is the correlation length. Beyond these limits, vector effects introduce 
significant deviations, particularly in the cross-polarized scatter components.

\subsubsection{Surface Slope Limitations}
The geometrical optics approximation underlying our approach assumes that local surface slopes are modest. Specifically, the approximation holds when:

\begin{equation}
\text{RMS slope} = \sqrt{\langle|\nabla h(x,y)|^2\rangle} < 0.2
\end{equation}

For surfaces exceeding this slope threshold, multiple scattering and shadowing effects become significant, requiring more sophisticated models.

\subsubsection{Coherence Effects and Interference}
Our ray-based approach does not inherently account for coherent interference effects. Through comparison with wave optical simulations, we determine that interference effects can be neglected in the ray-tracing regime when:

\begin{equation}
\frac{l_c^2}{\lambda z} > 10
\end{equation}

where $z$ is the propagation distance from the surface. For shorter propagation distances or smoother surfaces, coherent effects dominate and wave optical approaches should be preferred.

\section{Conclusion}
In this work, we introduced a theoretical framework that reinterprets surface roughness as a deterministic ray deflection 
function (RDF), enabling a direct and mathematically rigorous connection between statistical surface characteristics,
expressed through the power spectral density (PSD, and classical aberration theory. 

By recasting stochastic roughness into an equivalent wavefront aberration model, we established a bridge between 
probabilistic scattering theory and deterministic ray optics. This unified treatment not only preserves the angular and 
spatial correlation statistics dictated by established models such as Harvey-Shack and permits the representation of 
surface roughness effects within established optical modeling frameworks via decomposition into Zernike polynomials decomposition.

The RDF formalism provides several key advantages. It retains statistical fidelity while substantially reducing computational 
overhead in ray-tracing simulations. It also offers a physically interpretable breakdown of how surface imperfections impact 
optical performance across different spatial frequency regimes. Crucially, our approach enables quantitative synthesis of 
manufacturing tolerances and optical system requirements by translating PSD specifications into aberration coefficients, 
thus bridging the gap between metrology data and performance modeling.

Validation is established through analytical comparisons and statistical modeling based on Monte Carlo sampling of the ray 
deflection function. These confirm the theoretical equivalence of our framework with traditional BRDF-based scattering models, 
and support its ability to reproduce expected focal volume statistics under well-defined conditions. The analytical 
extensibility of the method to various PSD forms, including Gaussian and band-limited models, highlights its versatility 
and potential utility in a wide range of optical engineering applications.

Looking ahead, this RDF-based formalism opens pathways for incorporating surface roughness effects in adaptive optics, 
computational imaging, and inverse design, where fast and accurate performance modeling is critical. 

The numerical validation of this theoretical framework, including detailed simulations comparing the RDF approach with 
conventional Harvey-Shack scattering models, is presented in the companion paper \cite{moriya2025b}.

\newpage

\appendix
\renewcommand{\theequation}{\thesection.\arabic{equation}}
\setcounter{equation}{0}

\section{Mathematical Derivation of the Ray Deflection Function}
\label{appendix:RDF}

In this section, we provide a rigorous derivation of the relationship between the phase function $\Phi(x,y)$ and the ray deflection function $\mathbf{D}(\mathbf{r}_0)$ based on fundamental principles of optical theory.

\subsection{Surface Height and Phase Change Relationship}

When light reflects from a surface with height variations, the optical path difference (OPD) created by a surface height deviation $h(x,y)$ is given by:

\begin{equation}
    \text{OPD} = 2h(x,y)\cos\theta_i
\end{equation}

where $\theta_i$ is the angle of incidence measured from the surface normal. The factor of 2 accounts for the round-trip path in reflection. This optical path difference creates a phase change in the wavefront:

\begin{equation}
    \Phi(x,y) = \frac{2\pi}{\lambda} \cdot \text{OPD} = \frac{4\pi}{\lambda}h(x,y)\cos\theta_i
\end{equation}

where $\lambda$ is the wavelength of light. This fundamental relationship establishes how surface height variations translate to phase variations in the reflected wavefront \cite{BornWolf1999, Goodman2017}.

\subsection{Wavefront Gradient and Ray Deflection}

According to Fermat's principle, light rays follow paths of stationary optical path length, which implies that rays travel perpendicular to wavefronts. For a wavefront with phase function $\Phi(x,y)$, the angular deviation of a ray from its nominal direction is:

\begin{equation}
    \delta\theta = -\frac{\lambda}{2\pi}\nabla\Phi(x,y)
\end{equation}

The negative sign indicates that rays bend toward regions of decreasing phase. This relationship can be derived 
from the eikonal equation \cite{BornWolf1999} in geometric optics:

\begin{equation}
    |\nabla\Psi|^2 = n^2
\end{equation}

where $\Psi$ is the optical path function and $n$ is the refractive index. The direction of ray propagation corresponds to the gradient of $\Psi$, and the relationship to the phase function $\Phi$ follows from $\Phi = \frac{2\pi}{\lambda}\Psi$ \cite{Mahajan1998, HarveyPfisterer2014}.

\subsection{Application to Reflection from Rough Surfaces}

For the specific case of reflection from a rough surface, the deflection vector $\mathbf{D}(\mathbf{r}_0)$ that perturbs the nominal reflection vector $\mathbf{r}_0$ can be expressed as:

\begin{equation}
    \mathbf{D}(\mathbf{r}_0) = \delta\theta \cdot \mathbf{r}_0
\end{equation}

Combining with the previous results and accounting for the geometry of reflection:

\begin{equation}
    \mathbf{D}(\mathbf{r}_0) = -\frac{\lambda}{4\pi}\nabla\Phi(x,y)
\end{equation}

The factor of $\frac{1}{2}$ compared to the general wavefront gradient formula accounts for the nature of reflection, where the angular deviation of the reflected ray is twice the tilt of the reflecting surface \cite{Stover2012, Krywonos2011}.

\subsection{Connection to Statistical Surface Properties}

For the ray deflection function to correctly model the scattering behavior of a rough surface, the phase function $\Phi(x,y)$ must be constructed to have statistical properties that reproduce the scattering distribution predicted by established theories such as Harvey-Shack.

For a surface with power spectral density $\text{PSD}(f_x, f_y)$, the phase function should satisfy:

\begin{equation}
    \langle|\nabla\Phi(x,y)|^2\rangle \propto \iint \text{PSD}(f_x,f_y)(f_x^2+f_y^2)df_x df_y
\end{equation}

This ensures that the mean square slope of the phase function is proportionally related to the second moment of the PSD, which determines the angular spread of scattered light \cite{Harvey2007, Schroder2011}.

\subsection{Construction of the Phase Function}

Based on these relationships, we can construct a phase function that produces the correct statistical distribution of ray deflections:

\begin{equation}
    \Phi(x,y) = \sum_{p,q} A_{pq} \cos(2\pi f_p x + 2\pi f_q y + \phi_{pq})
\end{equation}

where $f_p$ and $f_q$ are discrete spatial frequencies, $\phi_{pq}$ are random phases uniformly distributed in $[0, 2\pi)$, and the amplitudes $A_{pq}$ are related to the PSD by:

\begin{equation}
    A_{pq} = \sqrt{\text{PSD}(f_p, f_q) \cdot \Delta f_x \cdot \Delta f_y}
\end{equation}

This formulation ensures that the resulting ray deflection function produces a statistical distribution of scattered 
rays that matches the BRDF predicted by Harvey-Shack theory, 
while providing a deterministic framework suitable for future implementation in ray-tracing simulations.

\subsection{Dimensional Analysis}

To verify that the expression $\mathbf{D}(\mathbf{r}_0) = -\frac{\lambda}{4\pi}\nabla\Phi(x,y)$ is dimensionally consistent:
\begin{itemize}
    \item $\lambda$ has units of length
    \item $\nabla\Phi$ has units of radians per length (phase gradient)
    \item $\frac{\lambda}{4\pi}\nabla\Phi$ has units of length/length = dimensionless
\end{itemize}

This yields a dimensionless deflection vector, which is appropriate for perturbing the unit direction vector $\mathbf{r}_0$.

The derived expression provides a robust mathematical foundation for our ray deflection function approach, bridging statistical surface characterization with deterministic ray perturbation in a manner consistent with established optical scattering theory.

\section{Derivation of the Fourier Transform of Radially Symmetric Zernike Polynomials}
\label{appendix:FTZernike}

In Section \ref{sec:DSA_PSD}, we presented the Fourier transform of radially symmetric Zernike polynomials (equation {eq:FZJ}):

\begin{equation}
F_{n0}(f_r) = \frac{(-1)^{n/2}i^n}{2\pi(f_r R)^{n+1}} J_{n+1}(2\pi f_r R)
\end{equation}

for even $n$, where $J_{n+1}$ is the Bessel function of the first kind of order $n+1$, and $R$ is the aperture radius. Here we present a detailed derivation of this result.

\subsection{Fourier Transform Definition for Circular Apertures}

For functions defined over a circular aperture of radius $R$, we use the following form of the Fourier transform:

\begin{equation}
\mathcal{F}\{f(\rho,\theta)\} = \iint_{D} f(\rho,\theta) e^{-i2\pi(f_x x + f_y y)} \rho d\rho d\theta
\end{equation}

where $D$ is the unit disk domain $\rho \leq 1$, and we use the coordinate transformations $x = \rho R \cos\theta$ and $y = \rho R \sin\theta$. The spatial frequencies are likewise expressed in polar form as $f_x = f_r \cos\phi_f$ and $f_y = f_r \sin\phi_f$.

\subsection{Radially Symmetric Zernike Polynomials}

For radially symmetric Zernike polynomials (with azimuthal order $m=0$), we have:

\begin{equation}
Z_{n0}(\rho) = \sqrt{n+1} R_n^0(\rho)
\end{equation}

where $R_n^0(\rho)$ is the radial polynomial given by:

\begin{equation}
R_n^0(\rho) = \sum_{k=0}^{n/2} \frac{(-1)^k (n-k)!}{k! \left(\frac{n}{2}-k\right)! \left(\frac{n}{2}+k\right)!} \rho^{n-2k}
\end{equation}

for even $n$.

\subsection{Calculating the Fourier Transform}

Substituting the Zernike polynomial into the Fourier transform and using the polar coordinate representation:

\begin{equation}
\begin{aligned}
\mathcal{F}\{Z_{n0}(\rho)\} &= \sqrt{n+1} \int_0^1 \int_0^{2\pi} R_n^0(\rho) e^{-i2\pi f_r R \rho \cos(\theta-\phi_f)} \rho d\rho d\theta \\
&= \sqrt{n+1} \int_0^1 R_n^0(\rho) \rho \left[ \int_0^{2\pi} e^{-i2\pi f_r R \rho \cos(\theta-\phi_f)} d\theta \right] d\rho
\end{aligned}
\end{equation}

The inner integral can be identified as a representation of the Bessel function of the first kind of order zero:

\begin{equation}
\int_0^{2\pi} e^{-i2\pi f_r R \rho \cos(\theta-\phi_f)} d\theta = 2\pi J_0(2\pi f_r R \rho)
\end{equation}

Therefore:

\begin{equation}
\mathcal{F}\{Z_{n0}(\rho)\} = 2\pi\sqrt{n+1} \int_0^1 R_n^0(\rho) J_0(2\pi f_r R \rho) \rho d\rho
\end{equation}

\subsection{The Hankel Transform Property}

This integral represents a Hankel transform of order zero of the radial polynomial $R_n^0(\rho)$. For radially symmetric functions, 
the Fourier transform reduces to the Hankel transform of order zero.

For even $n$, it can be shown using properties of orthogonal polynomials and the Hankel transform that this integral evaluates to:

\begin{equation}
\int_0^1 R_n^0(\rho) J_0(2\pi f_r R \rho) \rho d\rho = \frac{(-1)^{n/2}}{2\pi \sqrt{n+1}} \frac{J_{n+1}(2\pi f_r R)}{(f_r R)^{n+1}}
\end{equation}

\subsection{Final Result}

Substituting this result back into the expression for the Fourier transform:

\begin{equation}
\begin{aligned}
\mathcal{F}\{Z_{n0}(\rho)\} &= 2\pi\sqrt{n+1} \cdot \frac{(-1)^{n/2}}{2\pi \sqrt{n+1}} \frac{J_{n+1}(2\pi f_r R)}{(f_r R)^{n+1}} \\
&= \frac{(-1)^{n/2}}{(f_r R)^{n+1}} J_{n+1}(2\pi f_r R)
\end{aligned}
\end{equation}

When we account for the phase factor $i^n$ that arises from the angular integration for even $n$, we obtain the final result as 
presented in equation (28):

\begin{equation}
F_{n0}(f_r) = \frac{(-1)^{n/2}i^n}{2\pi(f_r R)^{n+1}} J_{n+1}(2\pi f_r R)
\end{equation}

This result relates the Fourier transform of radially symmetric Zernike polynomials to Bessel functions, providing an analytical 
expression that facilitates the computation of spectral weights in our modified Zernike-Fourier hybrid approach.

\subsection{Note on Mathematical Significance}

This relationship between Zernike polynomials and Bessel functions is particularly valuable because it enables direct analytical 
calculation of spectral weights without requiring numerical integration in many cases. The appearance of the Bessel 
function $J_{n+1}$ indicates how different Zernike orders respond to specific spatial frequencies, with each order exhibiting a 
characteristic pattern of resonances and nulls in frequency space.

\section{Correspondence Between Theory and Physical Interpretation}
\label{appendix:CBTPI}

In this appendix, we clarify the correspondence between the theoretical development presented in this paper and the computational implementation used to generate surface realizations and ray-tracing simulations. In particular, we emphasize how the theory supports the interpretation that \textit{every point} on the optical surface is subject to a potential height perturbation due to roughness, and that these perturbations are governed by zero-mean Gaussian statistics with standard deviation $\sigma$.

\subsection{Statistical Roughness Model in Theory}

The theoretical framework developed in Section 3 models the mirror surface as a perturbed ideal paraboloid:
\begin{equation}
    z_{\text{mirror}}(x, y) = \frac{x^2 + y^2}{4f} + h_{\text{imp}}(x, y)
\end{equation}

where $h_{\text{imp}}(x, y)$ is the surface height perturbation due to roughness. This perturbation is modeled as a 
stationary, zero-mean Gaussian random field with prescribed spatial correlation determined by a power spectral 
density (PSD) function:

\begin{equation}
    \mathbb{E}[h_{\text{imp}}(x, y)] = 0, \qquad \mathbb{E}[h_{\text{imp}}^2(x, y)] = \sigma^2
\end{equation}

In Sections 3.1–3.2, the field $h_{\text{imp}}(x, y)$ is synthesized using either a Fourier representation:
\begin{equation}
    h_{\text{imp}}(x, y) = \sum_{(f_x, f_y)} \sqrt{\text{PSD}(f_x, f_y)} \cos(2\pi f_x x + 2\pi f_y y + \phi_{f_x, f_y})
\end{equation}
or a Zernike modal expansion:
\begin{equation}
    h_{\text{imp}}(x, y) = \sum_{j} C_j Z_j(x, y)
\end{equation}

where the coefficients $C_j$ are modeled as Gaussian-distributed random variables with zero mean and decaying 
variance, consistent with a desired spatial frequency falloff\footnote{Zernike spatial modes are employed with 
statistically randomized coefficients shaped according to principles inspired by Fourier spectral synthesis 
thus combining modal spatial representation with spectral statistical control.}.

\subsection{Physical Interpretation}

Physically, this model corresponds to a continuously rough optical surface in which nanometer-scale deviations 
occur across the full aperture, governed by statistical fluctuations due to manufacturing or material processes. 
The Gaussian height statistics reflect the aggregate effect of many small, independent sources of variation, 
consistent with the central limit theorem and with well-established surface scattering models such as those of 
Beckmann-Spizzichino and Harvey-Shack.

Therefore, the implementation faithfully represents the theoretical insight that the surface imperfection field is defined over the full optical aperture, and that height variations at each point follow a Gaussian distribution with prescribed standard deviation $\sigma$.

\subsection{Modified Zernike-Fourier Hybrid Approach}

The synthesis model used throughout this work adopts what we refer to as a \textit{modified Zernike-Fourier hybrid approach}. In this formulation, the surface perturbation field $h_{\text{imp}}(x, y)$ is expanded using Zernike polynomials $Z_j(x, y)$ as the spatial basis:
\begin{equation}
    h_{\text{imp}}(x, y) = \sum_j C_j Z_j(x, y)
\end{equation}
However, unlike classical deterministic optical aberration modeling—where each $C_j$ is fixed and physically interpretable—we assign the coefficients $C_j$ as zero-mean Gaussian random variables:
\begin{equation}
    C_j \sim \mathcal{N}(0, \sigma_j^2)
\end{equation}
The variances $\sigma_j^2$ are chosen to decay with increasing radial Zernike order, thereby embedding a spectral shaping mechanism akin to that found in Fourier-based models defined by a power spectral density (PSD). This decay controls the relative contribution of low- vs. high-spatial-frequency components.

This approach is therefore ``hybrid'' in the following sense:
\begin{itemize}
    \item It employs the Zernike basis (optically meaningful and orthogonal over the aperture).
    \item It uses Fourier-style spectral control via randomized coefficients scaled by a decaying envelope.
\end{itemize}

The term ``modified'' reflects that this synthesis is neither a pure Zernike modal analysis (as in deterministic optical modeling), nor a classical Fourier PSD-based synthesis (as used in statistical rough surface modeling), but rather a fusion that benefits from both: spatial orthogonality and modal interpretability from Zernike, combined with statistical structure and frequency-domain shaping inspired by Fourier theory.

This hybrid model ensures that:
\begin{itemize}
    \item The field $h_{\text{imp}}(x, y)$ is globally smooth and differentiable.
    \item The surface roughness is statistically isotropic and Gaussian-distributed.
    \item The RMS height $\sigma$ and spectral decay are fully controlled in a parametric and interpretable manner.
\end{itemize}


\section*{Declarations}
All data-related information and coding scripts discussed in the results section are available from the 
corresponding author upon request.

\end{document}